\documentclass[pra,aps,twocolumn,superscriptaddress,showpacs]{revtex4}
\usepackage{psfig}

\begin{document}

\title{Conditional generation of arbitrary single-mode quantum states 
of light\\ by repeated photon subtractions}

\author{Jarom\'{\i}r Fiur\'{a}\v{s}ek}
\affiliation{Department of Optics, Palack\'{y} University, 
17. listopadu 50, 77200 Olomouc, Czech Republic}

\author{Ra\'{u}l Garc\'{\i}a-Patr\'on\,}
\affiliation{QUIC, Ecole Polytechnique, CP 165, 
Universit\'{e} Libre de Bruxelles, 1050 Brussels, Belgium }

\author{Nicolas J. Cerf\,}
\affiliation{QUIC, Ecole Polytechnique, CP 165, 
Universit\'{e} Libre de Bruxelles, 1050 Brussels, Belgium }

\begin{abstract}
We propose a scheme for the conditional generation of arbitrary finite
superpositions of (squeezed) Fock states in a single mode of a traveling
optical field. The suggested setup requires only a source of squeezed states, 
beam splitters, strong coherent beams, and photodetectors with single-photon sensitivity. The method does not require photodetectors with a high efficiency
nor with a single-photon resolution. 
\end{abstract}

\maketitle

\section{Introduction}

During recent years it has been widely recognized that 
non-classical states of light represent a valuable resource for
numerous applications ranging from ultra-high precision measurements 
\cite{Yurke86,Hillery93,Dowling98} to quantum lithography \cite{Boto00,Bjork01} 
and quantum information processing \cite{Bouwmeester}. 
It is often desirable to generate nonclassical states 
in traveling optical modes, as opposed to the cavity QED experiments 
where the generated state is confined in a cavity and can be probed
only indirectly. Many ingenious schemes have been 
proposed and experimentally demonstrated to generate the single-photon states 
\cite{Lvovsky01,Zavatta04} and various  multiphoton  entangled  states such as 
the GHZ states  \cite{Pan00}, cluster states \cite{Walther05}, 
and  the so-called NOON states 
\cite{Gerry02,Fiurasek02,Kok02,Mitchell04,Eisenberg04}.

Considerable attention has been also devoted to the
preparation of arbitrary single-mode states
\cite{Dakna99,Clausen01,Villas-Boas01,Zou04}  and, in particular,  the 
Schr\"{o}dinger cat-like superpositions of coherent states \cite{Dakna97,Lund04}
which can represent a valuable resource for quantum information processing
\cite{Jeong02,Ralph03}.  Experimental generation of arbitrary superpositions of 
vacuum and single-photon states has been accomplished using a parametric 
down-conversion with input signal mode prepared in a coherent state \cite{Resch02}, employing the quantum scissors scheme \cite{Pegg98,Babichev03} 
or conditioning on homodyne measurements on one part of a non-local single photon in two spatial modes \cite{Babichev04}. It is, however,  
very difficult to extend  these experiments to superpositions involving also
two or more photons. The known schemes for conditional generation of
arbitrary superpositions of Fock states require
single-photon sources and/or highly efficient detectors with single photon
resolution which represents a formidable experimental challenge.

In this paper, we propose a novel state preparation scheme, 
which does not require single-photon sources 
and can operate with high fidelity even with low-efficiency detectors that only 
distinguish the presence or absence of photons. Our scheme is
inspired by the proposal of Dakna \emph{et al.} \cite{Dakna99} 
who showed that an arbitrary single-mode 
state can be engineered starting from vacuum by applying a
sequence of displacements and single-photon additions. 
Our crucial observation is that if the initial state 
is a squeezed vacuum, then the single-photon addition can be replaced 
with single-photon subtraction \cite{Opatrny00,Kim04}, which is much more
practicable. Indeed, a single-photon subtraction can be achieved 
by diverting a tiny faction of the beam with a beam splitter 
towards a photodetector, so that a click means that a photon has been 
subtracted from the beam (this process becomes exact for a transmittance
tending to one). In fact, the single-photon subtraction from a squeezed vacuum 
has already been experimentally demonstrated \cite{Wenger04}, 
which provides a strong evidence for the practical feasibility of our scheme.
We note that the photon subtraction 
is an extremely useful tool that allows to generate 
states suitable for the tests of Bell inequality violation with balanced
homodyning \cite{Garcia04,Nha04}, can be used to improve the performance of
dense coding \cite{Kitagawa05}, and forms a crucial element 
of the  entanglement distillation schemes for continuous variables 
\cite{Browne03}.

The present paper is organized as follows.  
In Section~II, we explain the mechanism of state generation on the simplest non-trivial example of preparing a squeezed superposition 
of vacuum and single-photon states. Our setup then consists of 
two displacements and one conditional photon subtraction.  
We present the details of the calculation of the Wigner function of the generated state for a realistic setup involving imperfect photon
subtraction (obtained with imperfect detectors and beam splitters 
with a non-unity transmittance). 
In order to evaluate the performance of the scheme, we investigate  
the achieved fidelity and the preparation probability 
for various target states. 
In Section~III, we extend the scheme to the generation of an arbitrary
single-mode state and show how to calculate the displacements that need to be 
applied during the state preparation.  As an illustration, we consider 
the generation of several states which are squeezed superpositions of 
vacuum, single-photon, and two-photon Fock states. 
Finally, the conclusions are drawn in Section~IV. 

\vspace*{-9mm}

\section{Generation of a squeezed superposition of $|0\rangle$ and $|1\rangle$}

In this section, we introduce our setup for the generation of 
an arbitrary squeezed superposition of vacuum 
and single-photon state, which consists of two displacements 
with a photon subtraction in between, as schematically sketched in Fig. 1.
This setup represents a basic building block of our universal scheme:
as shown in Section~III, any squeezed superposition of the
first $N+1$ Fock states can be generated from a single-mode squeezed vacuum
by a displacement followed by a sequence of $N$ photon subtractions and
displacements.

\subsection{Pure-state description}

We first provide a simplified pure-state description of the setup,
assuming perfect detectors with single-photon resolution,
which will give us an insight into the mechanism of state generation. 
We will show that conditionally on observing a click of the photodetector PD,
the setup produces a squeezed superposition of vacuum and single-photon states, 
\begin{equation}
|\psi\rangle_{\mathrm{target}}= S(s) [c_0 |0\rangle+c_1 |1\rangle],
\label{target}
\end{equation}
where $S(s)=\exp[s(a^{\dagger 2}-a^2)/2]$ denotes the squeezing operator,
and $a$ ($a^\dagger$) is the annihilation (creation) operator.

Our state engineering procedure starts with a single-mode squeezed
vacuum state, which is generated in an optical parametric amplifier (OPA),
\begin{equation}
S(s_{\mathrm{in}})|0\rangle= \frac{1}{\sqrt{\cosh(s_{\mathrm{in}})}}\sum_{n=0}^\infty
\frac{\sqrt{(2n)!}}{2^n n!}[\tanh(s_{\mathrm{in}})]^n |2n\rangle,
\label{squeezing}
\end{equation}
where $s_{\mathrm{in}}$ denotes the initial squeezing constant.
The single-mode squeezed vacuum passes through three highly
transmitting beam splitters, which realize a sequence of a displacement
followed by a single-photon subtraction and another displacement.
The state is displaced by combining it on a highly unbalanced beam splitter
BS$_D$ with transmittance $T_D> 99$\% with a strong coherent state
$|\alpha /r_D\rangle$, 
where $r_D=\sqrt{R_D}$ and $R_D=1-T_D$ is the reflectance of BS$_D$ \cite{Paris96}. In the limit $T_D\rightarrow 1$,
the output beam is displaced by the amount $\alpha$. This method
has been used, e.g.,  in the continuous-variable quantum teleportation
experiments \cite{Furusawa98}. For the sake of simplicity, we shall assume 
that $T_D=1$ and the displacement operation is exact. 
The conditional single-photon subtraction requires
a highly unbalanced beam splitter BS with transmittance $T$, followed by
a photodetector PD placed on the auxiliary output port. A successful photon
subtraction is heralded by a click of the detector. In the limit
$T \rightarrow 1$, the most probable event leading to a click of the detector
is that exactly a single photon has been reflected from the beam splitter.
The probability of removing two or more photons is smaller by a factor of $1-T$
and becomes totally negligible in the limit $T \rightarrow 1$. The conditional
single-photon subtraction can be described by the non-unitary operator
\begin{equation}
X=t^n r\, a, 
\end{equation}
where  $n= a^\dagger a$ is the photon-number operator, while
$t=\sqrt{T}$ and $r=\sqrt{1-T}$ denote the amplitude
transmittance and reflectance of BS, respectively.

\begin{figure}[!t!]
\centerline{\psfig{figure=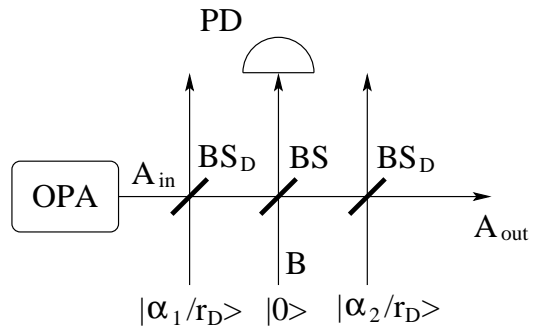,width=0.8\linewidth}}
\caption{Proposed experimental setup for generating 
$|\psi\rangle=S(s) [c_0 |0\rangle+c_1 |1\rangle]$. 
An optical parametric amplifier (OPA)
generates a single-mode squeezed vacuum state which then propagates through 
three highly unbalanced beam splitters BS$_D$, BS, and BS$_D$,
which realize a sequence of two displacements interspersed 
with one conditional photon subtraction. Successful state preparation
is heralded by a click of the photodetector PD.}
\end{figure}

The input-output transformation corresponding to the sequence of operations
in Fig. 1 reads
\begin{equation}
|\psi\rangle_{\mathrm{out}}=D(\alpha_{2}) X D(\alpha_{1}) S(s_{\mathrm{in}})|0\rangle,
\label{sequence}
\end{equation}
where $D(\alpha)=\exp(\alpha a^\dagger-\alpha^\ast a)$ is the displacement
operator.
We will show later on how the displacements $\alpha_{1}$ and $\alpha_{2}$
depend on the target state (\ref{target}), as well as how the initial squeezing
$s_{\mathrm{in}}$ depends on the target squeezing $s$ for a given 
transmittance $T<1$. But, to make it simple, let us
assume first that $T=1$ and $\alpha_1=-\alpha_2=\alpha$. Then, 
using $D(\alpha)^\dagger a D(\alpha) = a+\alpha$, the
conditionally generated state can be written as
\begin{equation}
|\psi\rangle_{\mathrm{out}}=(a+\alpha)S(s_{\mathrm{in}})|0\rangle . 
\end{equation}
Taking into account that $a$ and $a^\dagger$ transform under the squeezing operation 
according to
\begin{eqnarray}
S^\dagger(s) a S(s)= a \cosh(s)+a^\dagger \sinh(s), \nonumber \\
S^\dagger(s) a^\dagger S(s)= a^\dagger \cosh(s)+a \sinh(s),
\label{SaS}
\end{eqnarray}
we obtain 
\begin{eqnarray}
|\psi\rangle_{\mathrm{out}}&=&S(s_{\mathrm{in}})
[a\cosh(s_{\mathrm{in}})+a^\dagger\sinh(s_{\mathrm{in}})+\alpha]|0\rangle
\nonumber \\
&=&S(s_{\mathrm{in}})(\sinh(s_{\mathrm{in}})|1\rangle+\alpha|0\rangle).
\end{eqnarray}
We can see that by setting $\alpha=(c_0/c_1) \sinh(s_{\mathrm{in}})$,
we obtain the
target state (\ref{target}). This simple analysis illustrates the principle of
operation of the scheme shown in Fig. 1. However, the limit $T=1$ is unphysical,
because the probability of successful state generation vanishes 
when $T\rightarrow 1$. Let us now take into account $T<1$.

In order to simplify the expression (\ref{sequence}), we first rewrite all
displacement operators in a normally-ordered form,
$D(\alpha)=e^{-|\alpha|^2/2} e^{\alpha a^\dagger} e^{-\alpha^\ast a}$,
and we obtain
\begin{equation}
|\psi\rangle_{\mathrm{out}} 
\propto e^{\alpha_{2}a^\dagger} e^{-\alpha_{2}^\ast a} \, t^{n}\, a \,
e^{\alpha_{1}a^\dagger} e^{-\alpha_{1}^\ast a}S(s_{\mathrm{in}})|0\rangle.
\end{equation}
Next, we propagate the operator $t^n$ to the right by using the relations
\begin{eqnarray}
t^n e^{\alpha^\ast a}= e^{\alpha^\ast a/t } t^n, \qquad
t^n e^{\alpha a^\dagger}= e^{t\alpha a^\dagger} t^n.
\label{t}
\end{eqnarray}
After these algebraic manipulations we obtain
\begin{equation}
|\psi\rangle_{\mathrm{out}} \propto e^{\alpha_{2} a^\dagger} a \,
e^{t\alpha_{1} a^\dagger}  e^{-[\alpha_{2}^\ast+\alpha_{1}^\ast/t] a}  
\, t^{n}S(s_{\mathrm{in}})|0\rangle .
\end{equation}
Note that we have also moved to the right the operator 
$e^{-\alpha_{2}^\ast a}$ and used the fact that 
$e^{\alpha a^\dagger} e^{\beta^\ast a}=e^{-\alpha \beta^\ast} 
e^{\beta^\ast a} e^{\alpha a^\dagger}$.

The combined action of the operators $t^{n}S(s_{\mathrm{in}})$ on vacuum 
produces a single-mode squeezed vacuum state (\ref{squeezing}) just as 
without applying $t^{n}$, only 
with a lower squeezing constant $s$ satisfying
$\tanh(s)=t^{2}\tanh(s_{\mathrm{in}})$. Thus, we can write 
\begin{equation}
t^{n}S(s_{\mathrm{in}})|0\rangle \propto S(s)|0\rangle .
\end{equation}
Finally, we move the operator $e^{\alpha_2 a^\dagger}$ to the right,
using the formula
$e^{\alpha_2 a^\dagger}a=(a-\alpha_2)\, e^{\alpha_2 a^\dagger},$
which results in
\begin{equation}
|\psi\rangle_{\mathrm{out}} \propto (a-\alpha_{2})\, e^{\delta a^\dagger}
e^{-\gamma^\ast a} S(s) |0\rangle,
\end{equation}
where $\delta=\alpha_{2}+t\alpha_{1}$ and $\gamma=\alpha_{2}+\alpha_{1}/t$.
With the help of Eq.~(\ref{SaS}), we can write,
\begin{equation}
e^{\delta a^\dagger} e^{-\gamma^\ast a} S(s) |0\rangle
\propto
S(s)e^{[\delta\cosh(s)-\gamma^\ast\sinh(s)]a^{\dagger}} |0\rangle,
\end{equation}
which is a state with a generally non-zero coherent displacement.
This displacement vanishes provided that $\alpha_1$ and $\alpha_2$ 
satisfy
\begin{equation} 
(\alpha_2+t\alpha_1)\cosh(s)=(\alpha_2^\ast+\alpha_1^\ast/t)\sinh(s).
\label{condition}
\end{equation}
If the condition (\ref{condition}) holds, then the output reads
\begin{eqnarray}
|\psi\rangle_{\mathrm{out}} &\propto&
(a-\alpha_2) \, S(s) |0\rangle
\nonumber\\
&\propto&  S(s) [a^{\dagger}\sinh(s)-\alpha_2] |0\rangle.
\label{output}
\end{eqnarray}
where we have moved the squeezing operator to the left using Eq. (\ref{SaS}). 
The desired squeezed superposition of the first two Fock states (\ref{target}) 
can then be obtained by choosing
\begin{eqnarray} 
\alpha_2 &=& -\frac{c_{0}}{c_1} \sinh(s) ,\\
\alpha_1 &=& t\frac{[\tanh^2(s)-t^2]\alpha_2+(t^2-1)\tanh(s)
\alpha_2^\ast}{t^4-\tanh^2(s)},
\end{eqnarray}
where the displacement $\alpha_{1}$ is determined  from 
the condition (\ref{condition}).
Note that we may assume that the coefficient $c_{1}$ of the Fock state
$|1\rangle$ is non-zero; otherwise, no photon subtraction is needed
to generate the target state.

It should be emphasized that the generated state with our scheme
is a {\em squeezed} superposition of Fock states.
In principle, this squeezing could be removed by sending 
the conditionally generated state (\ref{output})
through a squeezer $S(-s)$. However, in many cases, the squeezing may not be an 
obstacle or may even represent an advantage, such as in the generation 
of Schr\"{o}dinger cat states $|\alpha\rangle-|-\alpha\rangle$ 
which can be for small
$|\alpha|$ very well approximated by a squeezed single-photon
state $S(s)|1\rangle$ \cite{Lund04,Kim04}.

\subsection{Realistic model}

We shall now present a more realistic description of the proposed scheme 
taking into account that the photodetectors exhibit only single-photon
sensitivity, but cannot resolve the number of photons in the mode, and have
a detection efficiency $\eta<1$. Such detectors have two outcomes, either a
click or a no-click. We model this detector as a sequence of a beam
splitter with transmittance $\eta$ followed by an idealized detector which
performs a projection onto the vacuum and the rest of the Hilbert space, 
$\Pi_0=|0\rangle\langle 0|$ (no click), $\Pi_1= \openone-|0\rangle\langle 0|$ 
(a click). 

Similarly as in Ref.~\cite{Garcia04}, it is convenient to work in the 
phase-space representation and consider the transformation 
of Wigner functions. The setup in Fig.~1 involves two modes, 
the principal mode A and an auxiliary mode B. 
The Gaussian Wigner function of the initial state of mode A after the first
displacement reads
\begin{equation}
W_G(r_A;\Gamma_A,d_A)=\frac{\sqrt{\det\Gamma_A}}{\pi}
e^{-(r_A-d_A)^T \Gamma_{A} (r_A-d_A)},
\label{Ginit}
\end{equation}
where $r_A=(x_A,p_A)^T$ is the vector of quadratures of mode A 
and  $d_A = z_1 \equiv \sqrt{2}(\Re \alpha_1,\Im\alpha_1)^T$ is the displacement.
The matrix $\Gamma_A$ is the inverse of the covariance matrix 
$\gamma_A$. Initially, the mode A 
is in a squeezed vacuum state and the covariance 
matrix is diagonal,  
$\gamma_A=\mathrm{diag}(e^{2s_{\mathrm{in}}},e^{-2s_{\mathrm{in}}})$.

In a second step, the modes A and B are mixed on an unbalanced 
beam splitter BS and then mode B subsequently passes through a (virtual) beam splitter of transmittance $\eta$ which models the imperfect detection with efficiency $\eta$. This transformation is a Gaussian completely positive (CP)
map $\mathcal{M}$, and the resulting state of modes A and B is still 
a Gaussian state with the Wigner function
\begin{equation}
\label{GaussianAB}
W_{AB}(r_{AB})=\frac{\sqrt{\det\Gamma_{AB}}}{\pi^2} 
 e^{-(r_{AB}-d_{AB})^T \Gamma_{AB}(r_{AB}-d_{AB})},
\end{equation}
where $r_{AB}=(x_A,p_A,x_B,p_B)^T$. The vector of the first moments 
$d_{AB}=(d_A,d_B)^T$ and the covariance matrix $\gamma_{AB}=\Gamma_{AB}^{-1}$
can be expressed in terms of the initial parameters of mode $A$ 
before the mixing on BS (i.e., $z_1$ and $\gamma_A$) as follows:
\begin{eqnarray}
d_{AB}&\equiv&\left( \begin{array}{c} d_{A} \\
d_B \end{array}\right)=S \left( \begin{array}{c} z_1 \\
0 \end{array}\right), \nonumber \\
\gamma_{AB}&=&S(\gamma_{A} \oplus I_{B})S^T +G, 
\label{CPmap}
\end{eqnarray}
where $S=S_{\eta}S_{BS}$, $S_{\eta}= I_A\oplus \sqrt{\eta}I_{B}$ and 
$ G=0_{A}\oplus (1-\eta)I_B $  model the inefficient photodetector, 
and $S_{BS}$ is a symplectic matrix which describes the coupling 
of the modes $A$ and $B$ on an unbalanced beam splitter BS,
\begin{equation}
S_{BS}=\left [\begin{array}{cccc}
t   & 0   & r  & 0   \\
0   & t   & 0  & r   \\
-r  & 0   & t  & 0   \\
0   & -r  & 0  & t  \\
\end{array}\right ].
\end{equation}

After the photon subtraction, the density matrix $\rho_{A,\mathrm{out}}$ 
of mode $A$ conditioned on a click of the photodetector PD measuring 
the auxiliary mode B becomes
\begin{equation}
\rho_{A,\mathrm{out}}=\mathrm{Tr}_B[\rho_{AB}
(\openone_A \otimes \Pi_{1,B})],
\label{photonsub}
\end{equation}
where $\mathrm{Tr}_B$ denotes a partial trace over mode B,
and $\rho_{AB}$ is the two-mode density matrix  
of the Gaussian state characterized by the Wigner function (\ref{GaussianAB}).
Then, after the second displacement of $z_2\equiv\sqrt{2}
(\Re \alpha_2,\Im\alpha_2)^T$, the Wigner function 
of mode A can be written as a linear combination of two Gaussian 
functions (\ref{Ginit}), namely
\begin{equation}
W(r)\, P =C_{1} W_{G}(r;\Gamma_{1},d_{1})+C_{2} W_{G}(r;\Gamma_{2},d_{2}),
\label{Wigner1}
\end{equation}
where $ P $ is the probability of successful generation of the target state. 
The expression (\ref{Wigner1}) can be derived by rewriting 
Eq.~(\ref{photonsub}) in the Wigner representation.  One uses the fact
that the Wigner function of the POVM element $\Pi_{1,B}$ is a difference 
of two Gaussian functions,
\begin{equation}
W_{\Pi_{1}}(r)=\frac{1}{2\pi}-\frac{1}{\pi} e^{-x^2-p^2},
\end{equation}
and that the trace of the product of two operators can be evaluated 
by integrating the product of their Wigner representations over the phase space.

\begin{figure}[!t!]
\centerline{\psfig{figure=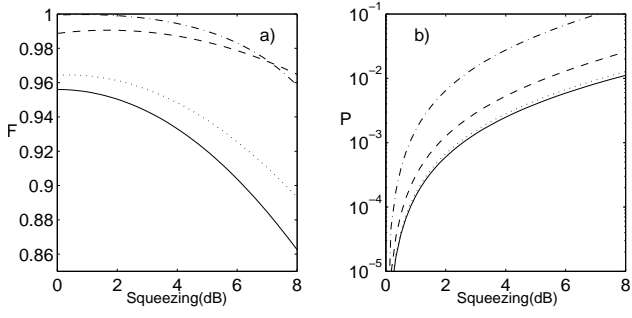,width=0.98\linewidth}}
\caption{
(a) Fidelity between the generated state and the
target state and (b) probability of successful generation 
as a function of the squeezing $s_{\mathrm{in}}$ for the four target states 
(\ref{targetone1}) (solid line),  (\ref{targetone2}) (dashed line), 
(\ref{targetone3}) (dot-dashed line) and  (\ref{targetone4}) (dotted line), 
with $T=0.95$ and $\eta=0.25$.}
\end{figure}

To define the matrices and vectors appearing in Eq.~(\ref{Wigner1}), 
we first divide 
the matrix $\Gamma_{AB}=\gamma_{AB}^{-1}$ into four sub-matrices with respect 
to the $A|B$ splitting,
\begin{equation}
\Gamma_{AB}=\left[ 
\begin{array}{cc}
\Upsilon_{A} & \sigma \\
\sigma^{T} & \Upsilon_{B}
\end{array}
\right].
\label{Gammadecomposition}
\end{equation}
The correlation matrix $\Gamma_1$ and the displacement $d_1$ appearing in 
the first term on the right-hand side of Eq.~(\ref{Wigner1}) are given by
\begin{eqnarray}
\Gamma_{1}&=&\Upsilon_{A}-\sigma\Upsilon_{B}^{-1}\sigma^T,
\nonumber \\
d_{1}&=&d_{A}+z_{2},
\nonumber \\
C_{1}&=&1.
\end{eqnarray}
Similarly, the formulas for the parameters of the second term read
\begin{eqnarray}
\Gamma_{2}&=&\Upsilon_{A}-\sigma\tilde{\Upsilon}_{B}^{-1}\sigma^T,
\nonumber \\
d_{2}&=&d_{A}+\Gamma_{2}^{-1}\sigma\tilde{\Upsilon}_{B}^{-1}d_{B}+z_{2},
\nonumber \\
C_{2}&=&-2
\sqrt{\frac{\det(\Gamma_{AB})}{\det(\Gamma_{2})\det(\tilde{\Upsilon}_{B})}}
\exp\left[-d_{B}^T M d_{B}\right],
\end{eqnarray}
where $\tilde{\Upsilon}_{B}=\Upsilon_{B}+I$ and 
\begin{equation}
M=\Upsilon_{B}\tilde{\Upsilon}_{B}^{-1}
-\tilde{\Upsilon}_{B}^{-1}\sigma^T \Gamma_{2}^{-1}\sigma \tilde{\Upsilon}_{B}^{-1}.
\end{equation}
Since all the Wigner functions in Eq. (\ref{Wigner1}) are normalized, 
the probability of a successful state generation can be calculated simply as the sum 
$P=C_1+C_2$.

\subsection{Examples}

In order to illustrate our method, let us consider the preparation of the  
following four squeezed superpositions of $|0\rangle$ and $|1\rangle$ states, 
\begin{eqnarray}
|\psi_{1}\rangle&=&S(s)|1\rangle,
\label{targetone1}\\
|\psi_{2}\rangle&=&S(s)\frac{1}{\sqrt{2}}(|0\rangle+ |1\rangle),
\label{targetone2}\\
|\psi_{3}\rangle&=&S(s)\frac{1}{\sqrt{10}}(3|0\rangle+ |1\rangle),
\label{targetone3}\\
|\psi_{4}\rangle&=&S(s)\frac{1}{\sqrt{10}}(|0\rangle+3 |1\rangle).
\label{targetone4}
\end{eqnarray}

\begin{figure}[!t!]
\centerline{\psfig{figure=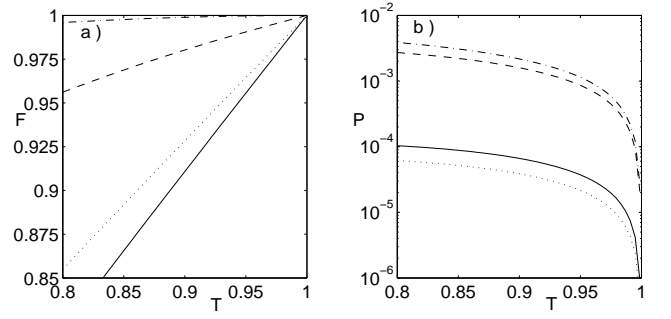,width=0.98\linewidth}}
\caption{
(a) Fidelity between the generated state and the
target state and (b) probability of successful generation as
a function of $T$ for the four target states 
(\ref{targetone1})--(\ref{targetone4}). 
The curves are plotted considering the optimal squeezing $s_{\mathrm{in}}$
for each state, namely $0.50$ dB for state (\ref{targetone1}) (solid line),
$1.66$ dB for state (\ref{targetone2}) (dashed line),
$0.85$ dB for state (\ref{targetone3}) (dot-dashed line), and 
$0.36$ dB for state (\ref{targetone4}) (dotted line). 
The curves are plotted for $\eta=0.25$.}
\end{figure}

The fidelity of the generated state for the target states 
(\ref{targetone1})--(\ref{targetone4}) is plotted in Fig.~2(a) 
as a function of squeezing. We can see that the conditionally prepared states 
are close to the desired states and their optimum fidelities are reached 
for a low initial squeezing (below $2$ dB), which is 
experimentally accessible. Although it is hardly visible in Fig~2(a),
there is typically a non-zero optimal value of the squeezing,
giving the highest fidelity. 
As shown in Fig.~2(b) the increase of the squeezing 
improves the probability of successful generation of the target state.
A comparison of Fig.~2(a) with Fig.~2(b) reveals a clear  
trade-off between the achievable fidelity and the preparation probability.

Figure~3(a) shows the  dependence of the fidelity on the beam splitter
transmittance $T$, considering the optimal squeezing for each of the states.
(Note that for state (\ref{targetone1}), we could not find numerically
the optimum squeezing, so we arbitrarily chose 
$s_{\mathrm{in}}=0.50$~dB as an optimal value in other to keep a reasonable
generation probability.)
We see that as $T$ approaches unity, the fidelity 
gets arbitrarily close to unity, while the probability of successful state 
generation $P \propto (1-T) \,\eta$ decreases, as shown in Fig. 3(b).
The value $T=0.95$ used in Fig.~2 seems to be a reasonable compromise between
the success rate ($P\approx 10^{-3}$ or $P\approx 10^{-4}$ depending on the target state) and the fidelity, $F\gtrsim 0.95$\%.

We also have studied the dependence of the fidelity on the detection efficiency 
$\eta$. The numerical results are shown in Fig.~4(a), where we can see that the
scheme is very robust in the sense that
the fidelity almost does not depend on $\eta$. 
Fidelities above $95$\%  could be reached even with $\eta$ of the order of a few
percent if $T$ is high enough. This is in agreement with the findings of 
Ref.~\cite{Garcia04}. However, a low $\eta$ reduces the preparation probability,
as shown in Fig.~4(b).

\begin{figure}[!t!]
\centerline{\psfig{figure=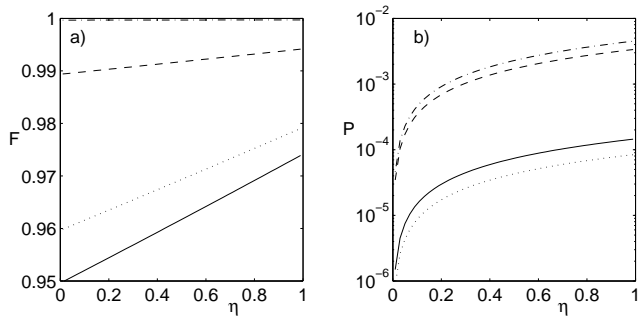,width=0.98\linewidth}}
\caption{
(a) Fidelity between the generated state and the
target state and (b) probability of successful generation as
a function of $\eta$ for the four target states 
(\ref{targetone1})--(\ref{targetone4}). 
The curves are plotted considering the optimal squeezing $s_{\mathrm{in}}$
for each state, namely $0.50$ dB for state (\ref{targetone1}) (solid line),
$1.66$ dB for state (\ref{targetone2}) (dashed line),
$0.85$ dB for state (\ref{targetone3}) (dot-dashed line), and 
$0.36$ dB for state (\ref{targetone4}) (dotted line). 
The curves are plotted for $T=0.95$.}
\end{figure}

\section{Arbitrary single-mode squeezed state}

In the preceding section, we have demonstrated that the combination
of two displacements and a photon subtraction allows us to build 
any squeezed superposition of $|0\rangle$ and $ |1\rangle $ states. 
In this section, we shall generalize this procedure to any squeezed
superposition of the first $N+1$ Fock states,
\begin{equation}
|\psi\rangle_{\mathrm{target}}= S(s)\sum_{n=0}^N c_n |n\rangle,
\label{targetN}
\end{equation}
and show that it can be prepared from a squeezed vacuum 
by applying a sequence of $N+1$ displacements 
interspersed with $N$ photon subtractions, as shown in Fig.~5.

\begin{figure}[!b!]
\centerline{\psfig{figure=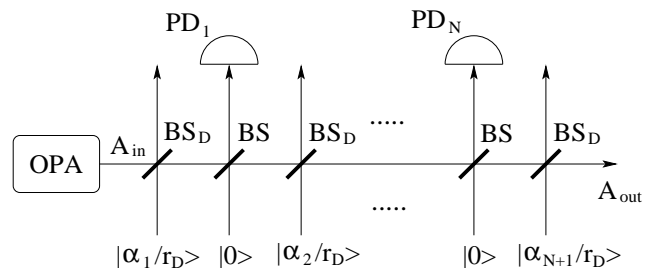,width=0.97\linewidth}}
\caption{Proposed experimental setup. An optical parametric amplifier (OPA)
generates a single-mode squeezed vacuum state which then propagates through 
 $2N+1$ highly unbalanced beam splitters BS$_D$ and BS, which realize 
 a sequence of $N+1$
 displacements interspersed with $N$ conditional photon subtractions. 
Successful state preparation is heralded by clicks of all $N$ photodetectors
 PD$_k$.}
\end{figure}

\subsection{Pure-state description}

As in the preceding section, we first provide 
a simplified pure-state description 
of the setup, assuming perfect detectors with single-photon resolution. 
This will allow us to determine the dependence of the
coherent displacements $\alpha_j$ on the target state (\ref{targetN}).
Generalizing the procedure presented in the preceding section,
the input-output transformation corresponding to the
sequence of operations in Fig.~5 reads
\begin{equation}
|\psi\rangle_{\mathrm{out}}=D(\alpha_{N+1}) X D(\alpha_{N}) X \ldots D(\alpha_2)X
D(\alpha_1) S(s_{\mathrm{in}})|0\rangle.
\end{equation}
In order to simplify this expression, we first rewrite all displacement 
operators in a normally-ordered form and then
move all the operators $t^n$ to the right using the relations (\ref{t}).
This results in the substitution $\alpha_j \rightarrow \alpha_j t^{N+1-j}$
and $\alpha_j^\ast \rightarrow \alpha_j^\ast t^{j-N-1}$ in the exponents.
Next, we propagate all the exponential operators
$e^{-t^{j-N-1}\alpha_{j}^\ast a}$ to the right,
\begin{equation}
|\psi\rangle_{\mathrm{out}} \propto e^{\alpha_{N+1}a^\dagger}  a
e^{t\alpha_{N}a^\dagger} a \ldots 
 a e^{t^N\alpha_{1}a^\dagger} e^{-\gamma^\ast a}  
 t^{Nn}S(s_{\mathrm{in}})|0\rangle,
\end{equation}
where $\gamma=\sum_{j=1}^{N+1}\alpha_j t^{j-N-1}$.
The  combined action of the operators $t^{Nn}S(s_{\mathrm{in}})$ on the vacuum produces a
single-mode squeezed vacuum state,
$t^{Nn}S(s_{\mathrm{in}})|0\rangle \propto S(s)|0\rangle$, where
$\tanh(s)=t^{2N}\tanh(s_{\mathrm{in}})$. After some algebraic manipulations we get
\begin{equation}
|\psi\rangle_{\mathrm{out}} \propto S(s)\prod_{j=1}^{N} 
(a \cosh(s)+a^\dagger \sinh(s)-\beta_j) |0\rangle,
\label{outputN}
\end{equation}
where 
\begin{equation}
\beta_j=\sum_{k=j+1}^{N+1}\alpha_{k} t^{N+1-k}.
\label{alphabeta}
\end{equation}
The formula (\ref{outputN}) is valid provided that the overall displacement
is zero. This corresponds to the constraint 
\begin{equation}
\cosh(s)\sum_{j=1}^{N+1} \alpha_{j}t^{N+1-j} =\sinh(s)
\sum_{j=1}^{N+1} \alpha_{j}^\ast t^{j-N-1} \, ,
\label{vanishingdisplacement}
\end{equation}
which generalizes condition~(\ref{condition}).

We now prove that an arbitrary superposition of the first $N+1$ Fock states 
$\sum_{n=0}^N c_n|n\rangle$ can be expressed as 
$\prod_{j=1}^N(A-\beta_j)|0\rangle \equiv\sum _{k=0}^N h_k A^k |0\rangle$, 
where $A=a\cosh(s)+a^{\dagger} \sinh(s)$ and 
$h_k$ are the coefficients of the characteristic polynomial whose roots are $\beta_j$. 
From the condition 
\begin{equation}
\sum_{k=0}^N h_{k}A^k |0\rangle = \sum_{n=0}^N c_n |n\rangle,
\label{matching}
\end{equation}
we can immediately determine the coefficients $h_N$ and $h_{N-1}$, because only 
the term $A^{N}$ gives rise to $a^{\dagger N}$ and, similarly, only the expansion of 
$A^{N-1}$ contains $a^{\dagger N-1}$. We thus get equations
\begin{equation}
h_{N}=\frac{c_N \sinh^{-N}(s)}{\sqrt{N!}}, \qquad 
h_{N-1}=\frac{c_{N-1} \sinh^{1-N}(s)}{\sqrt{(N-1)!}}.
\end{equation}
Once we know $h_{N}$ and $h_{N-1}$, we insert them back in Eq.~(\ref{matching}), and, from $\sum_{k=0}^{N-2} h_{k}A^k |0\rangle = \sum_{n=0}^N c_n |n\rangle
-(h_{N-1}A^{N-1}+h_N A^{N})|0\rangle$, we determine $h_{N-2}$ and $h_{N-3}$.
By repeating this procedure, we can find all coefficients $h_j$. This proves
that the condition (\ref{matching}) can be always met 
for any nonzero squeezing, hence our
method is indeed universal and allows us to generate {\em arbitrary}
superpositions. After finding the $h_j$'s, the coefficients $\beta_j$'s are calculated as the roots of the characteristic
polynomial $\sum_{k=0}^N h_k \beta^k$, and, finally, the $N+1$ displacements
$\alpha_{j}$'s are determined by solving the system of $N+1$ 
linear equations (\ref{alphabeta}) and (\ref{vanishingdisplacement}).

\subsection{Realistic model}
 
We shall now present a more realistic description of the proposed scheme,
which takes into account realistic photodetectors. 
After $k$-th photon subtraction and $k+1$-th displacement, the density 
matrix $\rho_{k,A}$  of mode $A$ conditioned on a click of photodetector measuring 
the auxiliary mode B is related to $\rho_{k-1,A}$ as follows,
\begin{equation}
\rho_{k,A}=D_{k+1}\mathrm{Tr}_B[\mathcal{M}(\rho_{k-1,A}\otimes |0\rangle_B\langle 0|)
(\openone_A \otimes \Pi_{1,B})]D^\dagger_{k+1},
\label{step}
\end{equation}
where $D_{k+1}=D(\alpha_{k+1})$ is a displacement operator and $\mathcal{M}$ denotes the Gaussian CP map (\ref{CPmap}) 
that describes mixing of the modes A and B on BS and accounts for imperfect detection. 
Since each step (\ref{step}) gives rise to a linear combination of two Gaussian states from a
Gaussian state, the Wigner function of the state $\rho_{k,A}$
can be written as a linear combination of $2^k$ Gaussian functions, 
\begin{equation}
W_{k}(r)P_{k}=\sum_{j=1}^{2^k} C_{j,k} W_{G}(r;\Gamma_{j,k},d_{j,k}),
\label{Wignerk}
\end{equation}
where $P_{k}$ is the probability of success of the first $k$ photon subtractions.
The correlation matrices $\Gamma_{j,k}$ and displacements $d_{j,k}$
after $k$ photon subtractions and $k+1$ displacements can be expressed in
terms of $\Gamma_{j,k-1}$ and $d_{j,k-1}$.

Similarly as in Section~IIB, 
we first define the real displacement vector $z_k \equiv \sqrt{2}(\Re
\alpha_k,\Im\alpha_k)^T$ and the two-mode covariance matrix and vector
of mean values after the action of the CP map $\mathcal{M}$,
\begin{eqnarray}
\left( \begin{array}{c} v_{j,k,A} \\
v_{j,k,B} \end{array}\right)&=&S \left( \begin{array}{c} d_{j,k} \\
0 \end{array}\right), \nonumber \\
\gamma_{j,k,AB}&=&S(\Gamma_{j,k}^{-1} \oplus I_{B})S^T +G.
\label{CPmapjk}
\end{eqnarray}
We also decompose the inverse matrix $\Gamma_{j,k,AB}=\gamma_{j,k,AB}^{-1}$ 
similarly as in~Eq. (\ref{Gammadecomposition}),
\begin{equation}
\Gamma_{j,k,AB}=\left[ 
\begin{array}{cc}
\Upsilon_{j,k,A} & \sigma_{j,k} \\
\sigma_{j,k}^{T} & \Upsilon_{j,k,B}
\end{array}
\right].
\label{Gammadecompositionjk}
\end{equation}
The $j$-th term in Eq. (\ref{Wignerk}) gives rise to
two new terms. The $(2j-1)$-th term is parametrized by 
\begin{eqnarray}
\Gamma_{2j-1,k}&=&\Upsilon_{j,k-1,A}-\sigma_{j,k-1}\Upsilon_{j,k-1,B}^{-1}\sigma_{j,k-1}^T,
\nonumber \\
d_{2j-1,k}&=&v_{j,k-1,A}+z_{k+1},
\nonumber \\
C_{2j-1,k}&=&C_{j,k-1}.
\end{eqnarray}
Similarly, the formulas for the $2j$-th term read
\begin{eqnarray}
\Gamma_{2j,k}&=&\Upsilon_{j,k-1,A}-\sigma_{j,k-1}\tilde{\Upsilon}_{j,k-1,B}^{-1}\sigma_{j,k-1}^T,
\nonumber \\
d_{2j,k}&=&v_{j,k-1,A}+\Gamma_{2j,k}^{-1}\sigma_{j,k-1}\tilde{\Upsilon}_{j,k-1,B}^{-1}v_{j,k-1,B}+z_{k+1},
\nonumber \\
C_{2j,k}&=&-2C_{j,k-1}
\sqrt{\frac{\det(\Gamma_{j,k-1,AB})}{\det(\Gamma_{2j,k})\det(\tilde{\Upsilon}_{j,k-1,B})}}
\nonumber \\
&&\times \exp\left[-v_{j,k-1,B}^T M v_{j,k-1,B}\right],
\end{eqnarray}
where $\tilde{\Upsilon}_{j,k-1,B}=\Upsilon_{j,k-1,B}+I$ and 
\begin{eqnarray*}
M&=&\Upsilon_{j,k-1,B}\tilde{\Upsilon}_{j,k-1,B}^{-1} \\
& &-\tilde{\Upsilon}_{j,k-1,B}^{-1}\sigma_{j,k-1}^T
\Gamma_{2j,k}^{-1}\sigma_{j,k-1}\tilde{\Upsilon}_{j,k-1,B}^{-1}.
\end{eqnarray*}
Using these formulas repeatedly starting from the initial ($k=0$) Gaussian
state~(\ref{Ginit}),
one obtains after $N$ iterations the Wigner function of the conditionally
generated state. The probability of state preparation can be calculated simply 
as the sum of $C_{j,N}$, $P=\sum_{j=1}^{2^N} C_{j,N}$.

\subsection{Examples}

We shall now consider, as an illustration, the generation of squeezed superpositions of $|0\rangle$, $|1\rangle$, and $|2\rangle$.
These states, namely, 
\begin{equation}
|\psi \rangle=S(s)\frac{1}{\sqrt{1+|c_{0}|^2+|c_{1}|^2}}(c_{0}|0\rangle + c_{1}|1\rangle + |2\rangle).
\label{targettwo}\\
\end{equation}
can be prepared with two photon subtractions. 
Here, we assume that the coefficient $c_{2}$ of the Fock state $|2\rangle$ is
non-zero (we arbitrarily take it equal to one). Otherwise, only one (or zero)
photon subtraction would be needed to generate the target state.
In the case of two photon subtractions interspersed with three displacements, 
Eq.~(\ref{outputN}) reduces to
\begin{eqnarray}
|\psi\rangle_{\mathrm{out}}&\propto & S(s)
[(\sinh(s)\cosh(s)+\beta_{1}\beta_{2})|0\rangle \\ \nonumber
 &-&(\beta_{1}+\beta_{2})\sinh(s)|1\rangle + \sqrt{2}\sinh^2(s) |2\rangle ].
\label{targettwovvv}
\end{eqnarray}
This state matches the target state~(\ref{targettwo}) if
\begin{equation}
\beta_{1,2}=\frac{-B\pm\sqrt{B^2-4C}}{2}  \, ,
\end{equation} 
where
\begin{eqnarray*}
B&=&\sqrt{2}\sinh(s) c_{1}, \\
C&=&\sqrt{2}\sinh ^2(s) c_{0}-\sinh(s)\cosh(s).
\end{eqnarray*} 
Equations~(\ref{alphabeta}) and (\ref{vanishingdisplacement}) 
allow us to calculate 
the displacements needed to generate this state. Assuming for simplicity 
that $c_0$, $c_1$ and $s$ are chosen such that $\beta_1$ and $\beta_2$ 
are both real, we obtain,
\begin{eqnarray} 
\alpha_3 &=& \beta_{2}, \nonumber \\ 
\alpha_2 &=& (\beta_{1}-\alpha_{3})/t,  \label{ambiguity1} \\
\alpha_1 &=& \frac{\tanh(s)(\alpha_{3}+\alpha_{2}/t)
-(\alpha_{3}+t\alpha_{2})}{t^2-\tanh(s)/t^2}  \nonumber.
\end{eqnarray}

\begin{figure}[!t!]
\centerline{\psfig{figure=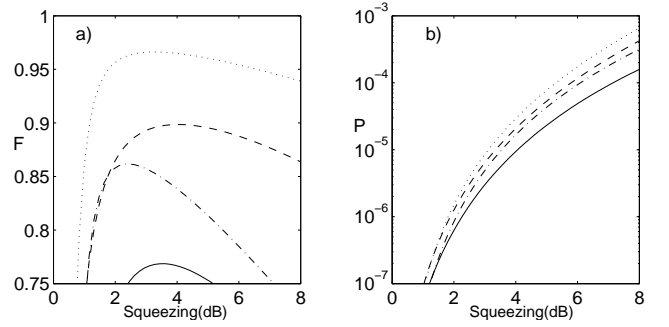,width=0.98\linewidth}}
\caption{
(a) Fidelity between the generated state and the
target state and (b) probability of successful generation 
as a function of the squeezing $s_{\mathrm{in}}$ for the four target states 
(\ref{targetwo1}) (solid line), 
(\ref{targetwo2}) (dashed line), (\ref{targetwo3}) (dot-dashed line), and 
(\ref{targetwo4}) (dotted line), with $T=0.95$ and $\eta=0.25$.}
\end{figure}

\begin{figure}[!t!]
\centerline{\psfig{figure=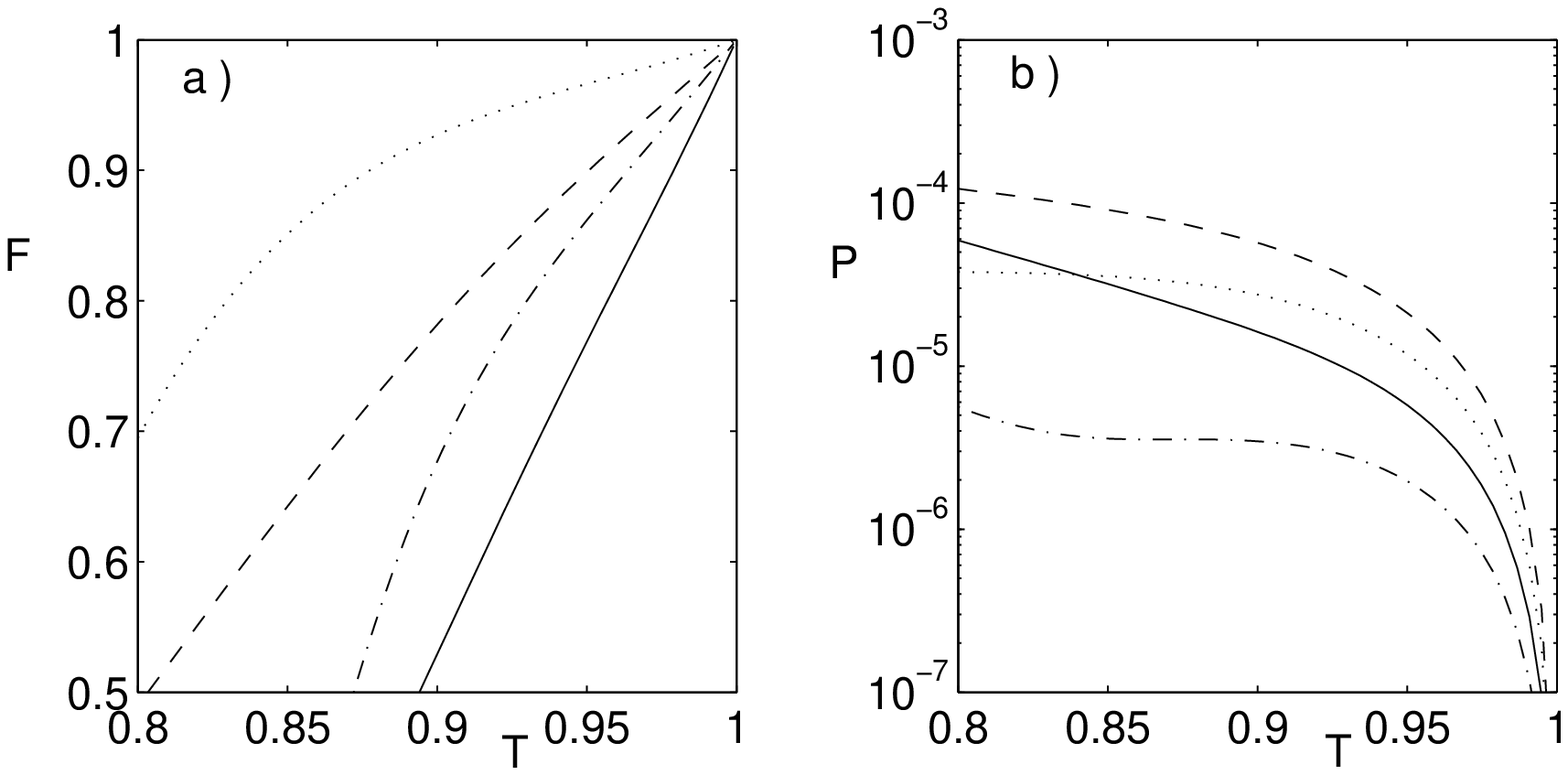,width=0.98\linewidth}}
\caption{
(a) Fidelity between the generated state and the
target state and (b) probability of successful generation as
a function of $T$ for the four target states 
(\ref{targetwo1})--(\ref{targetwo4}). 
The curves are plotted considering the optimal squeezing $s_{\mathrm{in}}$
for each state, namely
$3.54$ dB for state (\ref{targetwo1}) (solid line),
$4.02$ dB for state (\ref{targetwo2}) (dashed line),
$2.43$ dB for state (\ref{targetwo3}) (dot-dashed line), and 
$3.24$ dB for state (\ref{targetwo4}) (dotted line). 
The curves are plotted for $\eta=0.25$.}
\end{figure}

In order to illustrate our method, let us consider the
following four squeezed superpositions of the Fock states
$|0\rangle$, $|1\rangle$, and $|2\rangle$:
\begin{eqnarray}
|\psi_{1}\rangle&=&S(s)|2\rangle,
\label{targetwo1}\\
|\psi_{2}\rangle&=&S(s)\frac{1}{\sqrt{2}}(|1\rangle+ |2\rangle),
\label{targetwo2}\\
|\psi_{3}\rangle&=&S(s)\frac{1}{\sqrt{2}}(|0\rangle+ |2\rangle),
\label{targetwo3}\\
|\psi_{4}\rangle&=&S(s)\frac{1}{\sqrt{3}}(|0\rangle + |1\rangle + |2\rangle),
\label{targetwo4}
\end{eqnarray}
We plot the behavior of the fidelity and probability of generation of
the target states (\ref{targetwo1}) -- (\ref{targetwo4}) 
as a function of the squeezing (Fig.~6), beam-splitter transmittance (Fig.~7), and photodetector efficiency (Fig.~8). As in the preceding section,
we observe that the fidelity of the generation for any  
state gets arbitrarily close to one as $T$ approaches unity, 
as shown in Fig.~7. 
We also find that the fidelity is very robust against small detector efficiency 
$\eta$, as can be seen in Fig.~8. 
On the other hand, the preparation probability 
decreases with a growing $T$ and decreasing $\eta$,
as predicted by the equation $P \propto (1-T)^{2} \eta^{2}$.

All these features are very similar to those found in the preceding section, 
where we considered only states generated with one photon subtraction.
Let us now stress some new features. First, we note here the existence of 
a clear optimal squeezing, 
giving the maximum fidelity for each of the four studied states, 
see Fig.~6(a). 
Observing that the optimal squeezing has a higher value 
[from $2.4$ dB for state (\ref{targetwo3}) to $4$ dB for state
(\ref{targetwo2})] than those encountered in the case of one photon 
subtraction, we can expect an increasing value of 
the optimal squeezing for an increasing number of Fock states in the target superposition. It can be checked that the value of this optimal squeezing
tends to zero when $T$ tends to 100\%.

Another interesting fact is 
the existence of very different values of the optimum fidelity for 
different target states for a fixed $T=0.95$ and $\eta=0.25$, 
as shown in Fig.~6(a). For example, the squeezed 
two-photon state $S(s)|2\rangle$ is much more difficult to generate 
than the other three states (\ref{targetwo2})--(\ref{targetwo4}).
For the state $S(s)|2\rangle$, a transmittance of $T\gtrsim 0.99$ is 
necessary to reach a fidelity of $F\gtrsim 0.95$, resulting in a very low
probability of generation. This would make
the experimental generation of $S(s)|2\rangle$ with a good fidelity 
very challenging. In contrast, the squeezed balanced superposition state
(\ref{targetwo4}) can be generated with a high fidelity $F \gtrsim 0.90$
even with a transmittance $T \approx 0.90$.

\begin{figure}[!t!]
\centerline{\psfig{figure=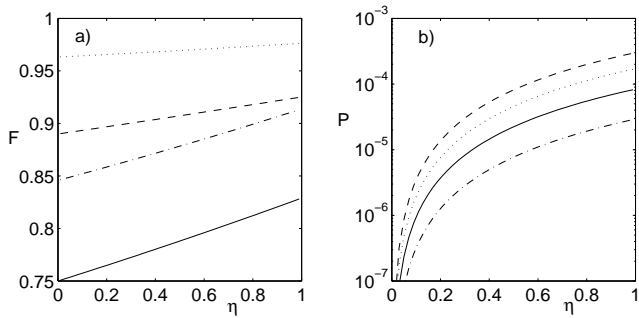,width=0.98\linewidth}}
\caption{
(a) Fidelity between the generated state and the
target state and (b) probability of successful generation as
a function of $\eta$ for the four target states (\ref{targetwo1})--(\ref{targetwo4}). 
The curves are plotted considering the optimal squeezing $s_{\mathrm{in}}$
for each state, namely
$3.54$ dB for state (\ref{targetwo1}) (solid line),
$4.02$ dB for state (\ref{targetwo2}) (dashed line),
$2.43$ dB for state (\ref{targetwo3}) (dot-dashed line), and 
$3.24$ dB for state (\ref{targetwo4}) (dotted line). 
The curves are plotted for $T=0.95$.}
\end{figure}

Another surprising fact arises when $\beta_1  \neq  \beta_2 $. 
Then, the equations (\ref{ambiguity1}) give two distinct 
sets of $\alpha_i$'s generating the same arbitrary target state, the second set
being obtained by making the exchange $\beta_1 \leftrightarrow \beta_2$. 
Considering the pure-state description and $T\rightarrow 1$,
the two alternative choices of displacements become strictly equivalent.
In contrast, when considering the realistic model with $T<1$,
these two solutions for the same target state do not have the exact same
behavior. As we can see in Fig.~9(a), one of the two solutions is indeed
more robust to decreasing $T$. However, the two solutions are rather 
similar as far as the probability of state generation is concerned,
as shown in Fig.~9(b).

\begin{figure}[!t!]
\centerline{\psfig{figure=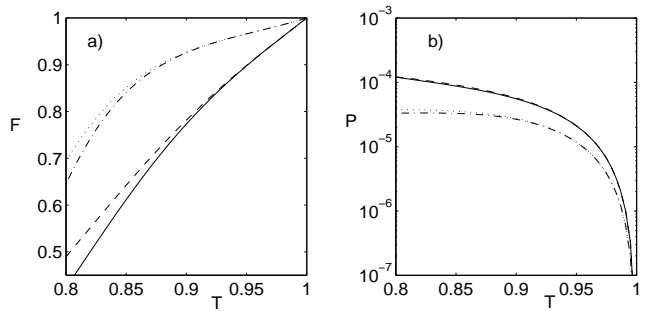,width=0.98\linewidth}}
\caption{
(a) Fidelity between the generated state and the
target state and (b) probability of successful generation as
a function of $T$ for the two target states (\ref{targetwo2}) 
and (\ref{targetwo4}).
The curves correspond to the two alternatives choices 
of the displacements $\alpha_1$ and $\alpha_2$ when considering
the optimal squeezing $s_{\mathrm{in}}$ for each state, namely $4.02$ dB for state~(\ref{targetwo2}) (dotted line, dot-dashed line), 
and $3.24$ dB for state~(\ref{targetwo4}) (dashed line, solid line). 
The curves are plotted for $\eta=0.25$.}
\end{figure}

\section{Conclusions}

In this paper, we have shown that an arbitrary single-mode state of light
can be engineered starting from a squeezed vacuum and applying a sequence of
displacements and single-photon subtractions. More precisely, the setup 
based on $N$ photon subtractions can be used to generate
a squeezed superposition 
of the $N+1$ first Fock states. If this remaining squeezing is not desired,
an extra squeezer can be added at the output of the setup in order 
to allow the generation of any (non-squeezed)
superposition of the $N+1$ first Fock states.

We have shown that the desired target state can be successfully produced
with arbitrarily high fidelity using a reasonably low squeezing
($ \lesssim 3 $dB) if the transmittance $T$ of the beam splitter
used for photon subtraction is sufficiently close to unity 
(e.g. $T\simeq 95$\%). This holds
even when inefficient photodetectors with single-photon sensitivity but
no single-photon resolution are employed, such as the commonly used avalanche
photodiodes. We have studied the dependence of
the achievable fidelity on the detection efficiency $\eta$,
and have found that the scheme is very robust in the sense 
that the fidelity does almost not depend on $\eta$. 
Fidelities above $95$\%  could be reached even with $\eta$ of the order of a few
percent if $T$ is high enough.
However, low $\eta$ and high $T$ drastically reduce the preparation probability,
so that a compromise has to be made when determining $T$.

Since our proposal does not require single-photon sources 
and can operate with low-efficiency photodetectors, its experimental implementation should be much easier than for the previous proposals,
in particular the one based on repeated photon additions~\cite{Dakna99}.
The recent demonstration of 
single-photon subtraction from a single-mode squeezed vacuum \cite{Wenger04}
provides a strong evidence for the practical feasibility of our scheme.
In this experiment,
a rather low overall detection efficiency ($\eta\approx 1$\%) was reported,
due to the spectral and spatial 
filtering of the beam measured by the photodetector. This is necessary because 
the OPA emits squeezed vacuum in several modes and the mode $A$ 
is selected in the
experiment by the strong local oscillator pulse in balanced homodyne detector.
Despite the filtering, the single-photon detector PD can be sometimes triggered 
by photons coming from other modes, so this would probably be the main source 
of imperfections in our scheme.

As a conclusion, we may reasonably assert that the 
generation of arbitrary squeezed superpositions of $|0\rangle$ and 
$|1\rangle$ should be experimentally achievable using our scheme
with the present technology, 
while some improvement of the filtering and detection efficiency would
probably be needed in order to extend the scheme to the preparation 
of squeezed superpositions of $|0\rangle$, $|1\rangle$, and $|2\rangle$.

\acknowledgments

We acknowledge financial support from the EU under projects 
COVAQIAL (FP6-511004) and RESQ (IST-2001-37559), from  
the Communaut\'e Fran\c{c}aise de Belgique under grant ARC 00/05-251, 
and from the IUAP programme of the Belgian
government under grant V-18. JF also acknowledges support under the Research 
project Measurement and Information in Optics MSM 6198959213 and  
from the grant  202/05/0498 of the Grant Agency of Czech Republic. 
R.G-P. acknowledges support from the Belgian foundation FRIA.

\end{document}